\documentclass[submitting]{nst}
\usepackage{subfigure,dcolumn}
\usepackage[T2A,T1]{fontenc}
\usepackage[russian,english]{babel}
\usepackage{multirow}
\usepackage{listings}
\begin{document}

\title{Contour-informed inter-patient deformable registration of Head-and-Neck patients}

\author{Xingyue Ruan}
\affiliation{Center for Proton Therapy, Paul Scherrer Institut, Villigen-PSI, Switzerland}
\affiliation{Shanghai Institute of Applied Physics, Chinese Academy of Sciences, Shanghai, China}

\author{Xia Li}
\affiliation{Center for Proton Therapy, Paul Scherrer Institut, Villigen-PSI, Switzerland}
\affiliation{Department of Computer Science, ETH Zurich, Zurich, Switzerland}

\author{Muheng Li}
\affiliation{Center for Proton Therapy, Paul Scherrer Institut, Villigen-PSI, Switzerland}
\affiliation{Department of Physics, ETH Zurich, Zurich, Switzerland}

\author{Barbara Bachtiary}
\affiliation{Center for Proton Therapy, Paul Scherrer Institut, Villigen-PSI, Switzerland}

\author{Antony Lomax}

\affiliation{Center for Proton Therapy, Paul Scherrer Institut, Villigen-PSI, Switzerland}
\affiliation{Department of Physics, ETH Zurich, Zurich, Switzerland}

\author{Zhiling Chen}
\affiliation{Shanghai Advanced Research Institute, Chinese Academy of Sciences, Shanghai, China}

\author{Ye Zhang}
\email[Corresponding author, ]{E-mail address: ye.zhang@psi.ch}

\affiliation{Center for Proton Therapy, Paul Scherrer Institut, Villigen-PSI, Switzerland}

\begin{abstract}
\textbf{Background and Purpose:} Voxel-based analysis (VBA) helps to identify dose-sensitive regions by aligning individual dose distributions within a common coordinate system (CCS). Accurate deformable image registration (DIR) is essential for addressing anatomical variability across patients. To improve both global and region-specific alignment, we enhanced our in-house DIR algorithm (CPT-DIR) by incorporating contour-informed regularisations. We tested its performance specifically for head-and-neck (HN) CT images.\\
\hspace*{1em} \textbf{Materials and Methods}:  We developed and evaluated contour-informed CPT-DIR using CT images from 37 HN patients, including seven with ground-truth dose distributions for dose warping validation. Bone contours were automatically generated using TotalSegmentator, while other organs at risk (OARs) were manually delineated. Contour-based constraints, such as Dice Similarity, were integrated to enhance registration in clinically relevant regions. The global registration results were evaluated using MAE, SSIM and PSNR. Geometric accuracy and warped dose accuracy were assessed using Dice Similarity Coefficient (DSC) and Dose-Organ Overlap (DOO). CPT-DIR with and without constraints performance was benchmarked against conventional B-spline.\\
\hspace*{1em} \textbf{Results:} CPT-DIR achieved superior accuracy with a MAE of 98.9 ± 6.3 HU, lower than 179.1 ± 17.8 HU for B-spline. Incorporating brainstem contours as regularisation improved the DSC from 0.604 ± 0.116 to 0.878 ± 0.017 and DOO from 0.430 ± 0.117 to 0.753 ± 0.043 for brainstem. Across all metrics, the enhanced CPT-DIR outperformed the B-spline, confirming its advantages in geometric accuracy and dosimetric consistency.\\
\hspace*{1em} \textbf{Conclusions:} The integration of contour-informed regularisation in CPT-DIR improved DIR accuracy, particularly in anatomically and dosimetrically relevant regions. This enhanced spatial alignment enabled more precise dose mapping for VBA and demonstrated strong potential for advancing reliable inter-patient dosimetric studies in HN radiotherapy.
\end{abstract}


\maketitle

\section{Introduction}

The Dose Volume Histogram (DVH) is a widely used tool in radiotherapy for quantifying and assessing dose distributions within target volumes and organs at risk (OARs). Condensing complex three-dimensional dose information into a single, interpretable graph, known as a DVH, offered a practical statistical overview of dose coverage and homogeneity within specified regions~\cite{bib:1,bib:2,bib:3}. However, DVHs reduce the 3D dose distributions in anatomical structures to 1D curves, losing all spatial information, and growing evidence indicates that different subregions within the same organ can exhibit varying levels of radiosensitivity and response~\cite{bib:4,bib:5,bib:6,bib:7}. Furthermore, the delineation of target volumes and OARs is subject to contouring uncertainties, which can compromise the accuracy and reliability of DVH-based evaluation~\cite{bib:8}. To address these limitations and gain deeper insights into patient-specific responses to radiation, Voxel-Based Analysis (VBA) has emerged as a promising alternative~\cite{bib:1,bib:6}. By assessing dose-outcome relationships at the voxel level, VBA enabled a more detailed and spatially resolved understanding of radiation effects, overcoming the averaging limitations inherent in traditional DVH approaches and removing the prior definition~\cite{bib:9}. To investigate spatially resolved dose-response relationships in VBA, individual patient images must first be normalised to a reference anatomy, so-called the common coordinate system (CCS). This was achieved by applying deformation vector fields (DVFs) derived from deformable image registration (DIR) to warp each patient’s dose distribution onto the CCS. Statistical analyses were then performed to identify voxels with significant correlations to specific toxicities, which were grouped into sub-regions of interest. These regions could be prioritised for sparing in future treatment planning or targeted for further research~\cite{bib:5}. By identifying critical dose-sensitive areas, VBA offered a robust framework for optimising radiotherapy plans and advancing personalised treatment strategies~\cite{bib:5,bib:10,bib:11,bib:12,bib:13}.\\
\hspace*{1em} As the foundation of VBA, accurate spatial normalisation via DIR was essential to ensure reliable and meaningful outcomes~\cite{bib:1,bib:6,bib:14}. Classical DIR methods, such as B-spline-based algorithms, were commonly used in VBA because they can model complex deformations efficiently~\cite{bib:15}. However, these traditional algorithms were primarily optimised for global alignment and often fail to achieve precise region-specific registration while maintaining overall anatomical consistency~\cite{bib:16,bib:17,bib:18}. This posed a significant challenge when aligning critical structures, such as organs-at-risk and tumour sub-regions, where accurate dose mapping was critical for voxel-level analysis. Accurate DIR remained a considerable challenge, particularly in the presence of substantial inter-patient anatomical variability and low local image contrast. Significant morphological differences, organ deformations, and treatment-induced changes could lead to severe registration errors, resulting in misaligned dose distributions and unreliable voxel-wise dose-response correlations~\cite{bib:19}. Therefore, enhancing the robustness and precision of DIR was essential to improve the reliability of VBA and fully realise its potential in radiotherapy outcome modelling. Building on our previous CPT-DIR (Continuous sPatial and Temporal-Deformable Image Registration) framework, which employed implicit neural representation to generate precise and continuous velocity vector field (VVF) for 4D medical images interpolation and generation~\cite{bib:20}, we presented here an enhanced version designed for VBA applications with large inter-patient anatomical variability.  To improve both global and regional registration accuracy, strengthened image alignment, and enhanced the reliability of dose warping and VBA, the proposed contour-informed deformable image registration approach integrated contour-informed regularisations to enhance spatial normalization of images for voxel-based analysis in HN radiotherapy.

\section{Materials and Methods}

\subsection{Dataset and template selection}
\begin{figure*}[h]
\includegraphics
  [width=1\hsize]
  {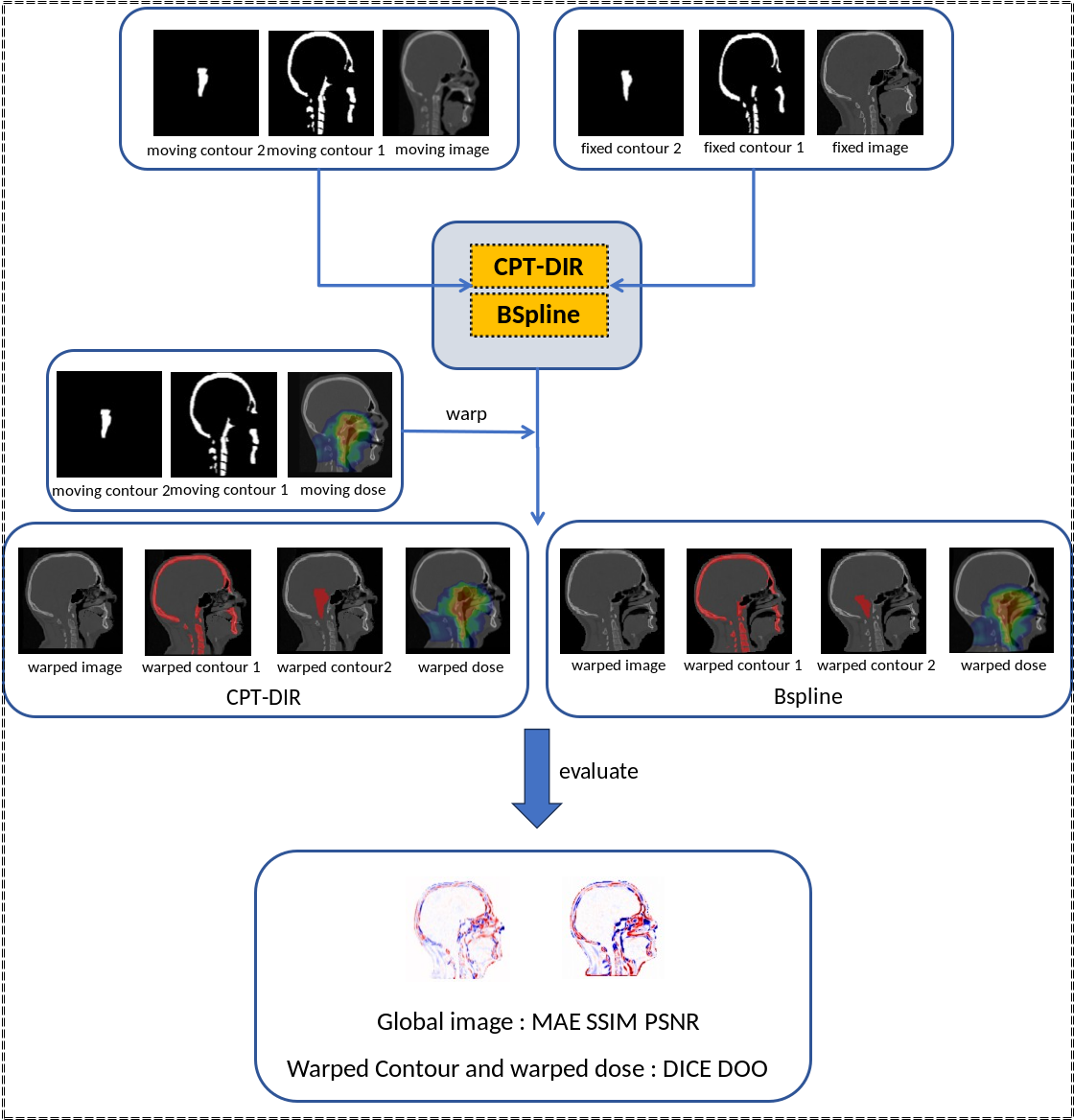}
\caption{Overall framework of this study. CT images and bone contours served as default inputs, while brainstem contours were optionally included for a subset of seven patients to evaluate OAR contour registration accuracy in the resulting warped images.}
\label{fig:1}
\end{figure*}

This study used anonymised CT images from 37 HN cancer patients treated at the Center for Proton Therapy, Paul Scherrer Institute (PSI-CPT). For 7 selected patients, intensity-modulated proton therapy (IMPT) plans were generated using the in-house treatment planning system (FIONA) as a proof-of-concept for testing the proposed DIR performance. All CT scans had a spatial resolution of 0.97 mm × 0.97 mm × 2 mm. To enable accurate registration comparisons, each image was manually cropped to include only anatomical structures above the sixth cervical vertebra (C6), as shown in Figure~\ref{fig:1}. Bone structures were automatically segmented using TotalSegmentator~\cite{bib:21}, while OARs, including the brainstem, eyes, optic nerves, and temporal lobes, were manually contoured by experienced specialists in radiological imaging. Inter-patient similarity was quantified as the mean Structural Similarity Index Measure (SSIM) between each image and all other images in the dataset, with values ranging from 0.797 to 0.909. Two CT images with an average SSIM greater than 0.9 were randomly selected as CCS1 and CCS2. This study adhered to ethical standards for research involving human data. Informed consent was obtained from all patients for the use of their anonymized data in scientific research.

\subsection{Contour-informed regularisations for upgraded CPT-DIR}
This study employed CPT-DIR~\cite{bib:19} as the core framework for image registration, which enabled efficient and accurate alignment by leveraging a VVF to generate the DVF. The VVF represented voxel-wise motion velocities over time, allowing for precise and continuous deformation due to its intrinsic reversibility. The DVF was obtained through temporal integration of the VVF, ensuring smooth and anatomically consistent registration across images. More specifically, the forward DVF ($\varphi_{0 \to 1}$) mapped the moving image ($I_{1}$) to the fixed image ($I_{0}$), while the backwards field ($\varphi_{1 \to 0}$) performed the inverse, ensuring bidirectional consistency and improved robustness. By incorporating spatiotemporal modelling, this framework effectively mitigated registration errors arising from substantial anatomical differences across patients, offering a novel solution for high-precision alignment in complex anatomical regions. In the original CPT-DIR framework, the loss function was defined as:

\begin{equation}\label{eq:1}
\begin{split}
	L_0 = D(I_0, I_1 \circ \varphi_{0 \to 1})+\lambda_{1}R_{1}(\varphi_{0 \to 1})+\lambda_{2}\sum\limits_{t=0}^{t=1}R_{2}(\upsilon_{t})+\\\lambda_{3}R({\omega})+D(I_1, I_0 \circ \varphi_{1 \to 0})+\lambda_{1}R_{1}(\varphi_{1 \to 0})
\end{split}
\end{equation}
where $\lambda_{i}$ was the weight of each term, and\\

$\bullet$\quad $D(I_0, I_1 \circ \varphi_{0 \to 1})$ represented the image similarity metric, which measured the difference between the reference image $I_{0}$ and the warped moving image $I_1 \circ \varphi_{0 \to 1}$ using DVF $\varphi_{0 \to 1}$. $D(I_1, I_0 \circ \varphi_{1 \to 0})$ was analogous to $D(I_0, I_1 \circ \varphi_{0 \to 1})$.\\

$\bullet$\quad $R_{1}(\varphi_{0 \to 1})$ and $R_{1}(\varphi_{1 \to 0})$ were regularisation terms applied to DVF $\varphi_{0 \to 1}$ and DVF $\varphi_{1 \to 0}$ for enforcing smoothness and physical plausibility.\\

$\bullet$\quad $\sum\limits_{t=0}^{t=1}R_{2}(\upsilon_{t})$ was a regularisation term applied to the VVF $\upsilon_{t}$, which controled gradient smoothness and ensures coherent motion trajectories.\\

$\bullet$\quad $R({\omega})$ was a regularisation term for the network parameters $\omega$, avoiding the numerical accuracy sacrifice from trilinear interpolation.\\

To extend this approach to be ‘contour-informed’, we incorporated the Dice Similarity metric into the original optimisation objective to improve local alignment between corresponding anatomical contours. This metric, defined in Equation (2), measured spatial overlap between predicted and reference contours (X and Y), guiding contour alignment during optimisation. The final loss function, shown in Equation (3), promoted consistent registration in anatomically relevant regions.
\begin{equation}\label{eq:2}
\text{DSC}(X,Y) = 2 \cdot \frac{|X \cap Y|}{|X| + |Y|}
\end{equation}
\begin{equation}\label{eq:3}
L = \alpha_{0} L_{0} + \sum\limits_{i=1}^{n} \alpha_{i} \cdot {DSC}(\text{Contour}_{i}(fixed),\text{Contour}_{i}(warped)
\end{equation}

where $\alpha$ were weighting coefficients, balancing the optimisation weights of the Dice Similarity metric relative to other components in optimisation processing. Higher weight coefficients were assigned to corresponding anatomical contours, enforcing stricter alignment constraints for these critical structures during the training process, so the weight coefficients were assigned according to the magnitude of the number of image voxels of corresponding contours.

\subsection{Evaluation of DIR performance for VBA}
The enhanced CPT-DIR algorithm took CT images and corresponding bone contours as default inputs. Each CT image from 36 non-CCS patients was used as the moving image and respectively registered to the selected CCS Figure~\ref{fig:1}. To evaluate the overall registration accuracy, three quantitative metrics were used:\\
\hspace*{2em}$\bullet$\quad Mean Absolute Error (MAE): Measures voxel-wise intensity differences between warped CTs and the CCS.\\
\hspace*{2em}$\bullet$\quad Structural Similarity Index Measure (SSIM): Evaluates structural consistency between contours of warped CTs and CCS.\\
\hspace*{2em}$\bullet$\quad Peak Signal-to-Noise Ratio (PSNR): Assesses image fidelity and registration quality.\\
The registration was performed and compared with and without the additional constraint of the brainstem to investigate the effect of additional contour-informed features. The resulting DVFs were used to warp all OAR contours and dose distributions to the CCS, allowing for a comprehensive evaluation of inter-patient registration accuracy and its associated dosimetric impacts. Warped contours and dose maps were further assessed using:\\
\hspace*{2em}$\bullet$\quad Dice Similarity Coefficient (DSC): Measures spatial overlap between warped contours and reference structures in the CCS.\\
\hspace*{2em}$\bullet$\quad Dose-Organ Overlap (DOO) Score~\cite{bib:22}: Quantifies the overlap between warped dose distributions and OARs to assess dose warping accuracy.\\
\\
\hspace*{1em}DSC could directly reflect the accuracy of contour registration of images, while DOO, on the basis of DSC, reflected the alignment degree of doses. Similar to DSC, the value range of DOO also belonged to 0 to 1, where 1 indicated that the registered doses were completely consistent, and 0 indicated that the registered doses were completely inconsistent.  In addition, Hausdorff Distance95 (HD95) also was utilized to evaluate accuracy of warped brainstem contour. CPT-DIR performance was benchmarked against conventional B-spline DIR, which served as the baseline for all comparisons~\cite{bib:23}.\\
\hspace*{1em}T-test statistical analysis was employed to compare the quantitative outcomes (DSC and DOO) across different DIR methods (B-spline, CPT-DIR(wo), and CPT-DIR(w)), which was used to determine the statistical significance of improvements in these metrics. All statistical analyses were performed using with a significance level set at $\alpha$= 0.05.\\
\hspace*{1em}In this study, the MAE, SSIM, and PSNR were computed using the scikit-image package, while the DSC and DOO metrics were implemented through custom-developed algorithms. $\alpha$ was set to 1 for $L_{0}$, 0.1 for the bone contour, and 0.01 for the brainstem. The results of CPT-DIR with bone and brainstem contour optimization were CPT-DIR(w), the results of CPT-DIR with bone but without brainstem contour optimization were CPT-DIR(wo). For the B-spline image registration, the following parameters were employed: a maximum iteration count ({max\_its}) of 100, a control point grid spacing ({grid\_spac}) of 20 × 20 × 20 $mm^{3}$, a regularization weight (regularization\_lambda) of 0.005, and a multi-resolution level (res) of 4 × 4 × 2. The CT images of all non-CCS patients were registered to two selected CCSs (denoted as CCS1 and CCS2) using both the CPT-DIR and B-spline algorithms.\\
\hspace*{1em}This study was conducted based on Python, all experiments ran on an AMD EPYC 7317P 3.0GHz 16-core CPU and the network was trained and inferenced on a NVIDIA RTX 4090.

\section{Results}
\begin{figure*}[h]
\includegraphics
  [width=1\hsize]
  {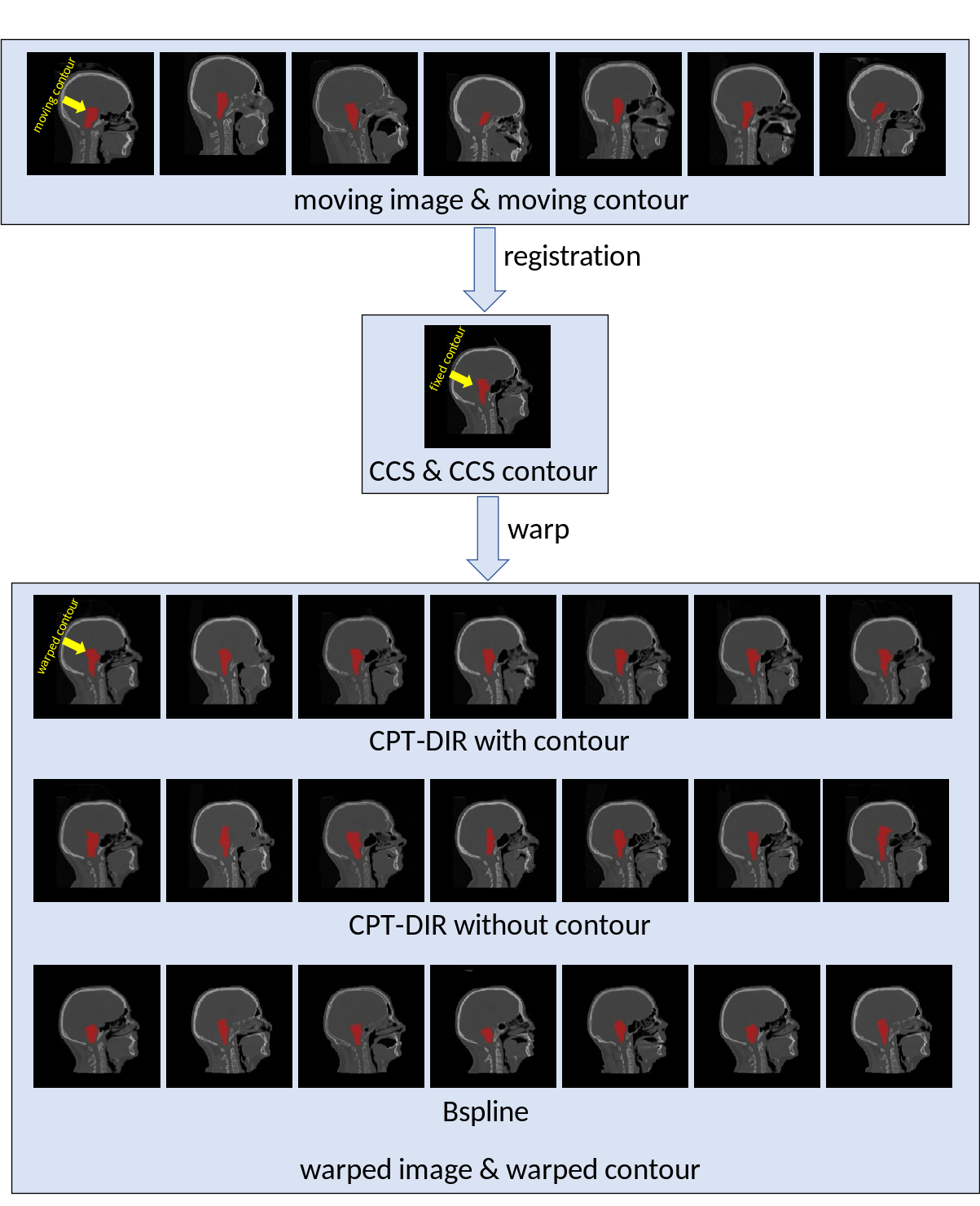}
\caption{Comparison of warped images and corresponding warped brainstem contours for CCS1.}
\label{fig:2}
\end{figure*}

\begin{figure*}[h]
\includegraphics
  [width=1\hsize]
  {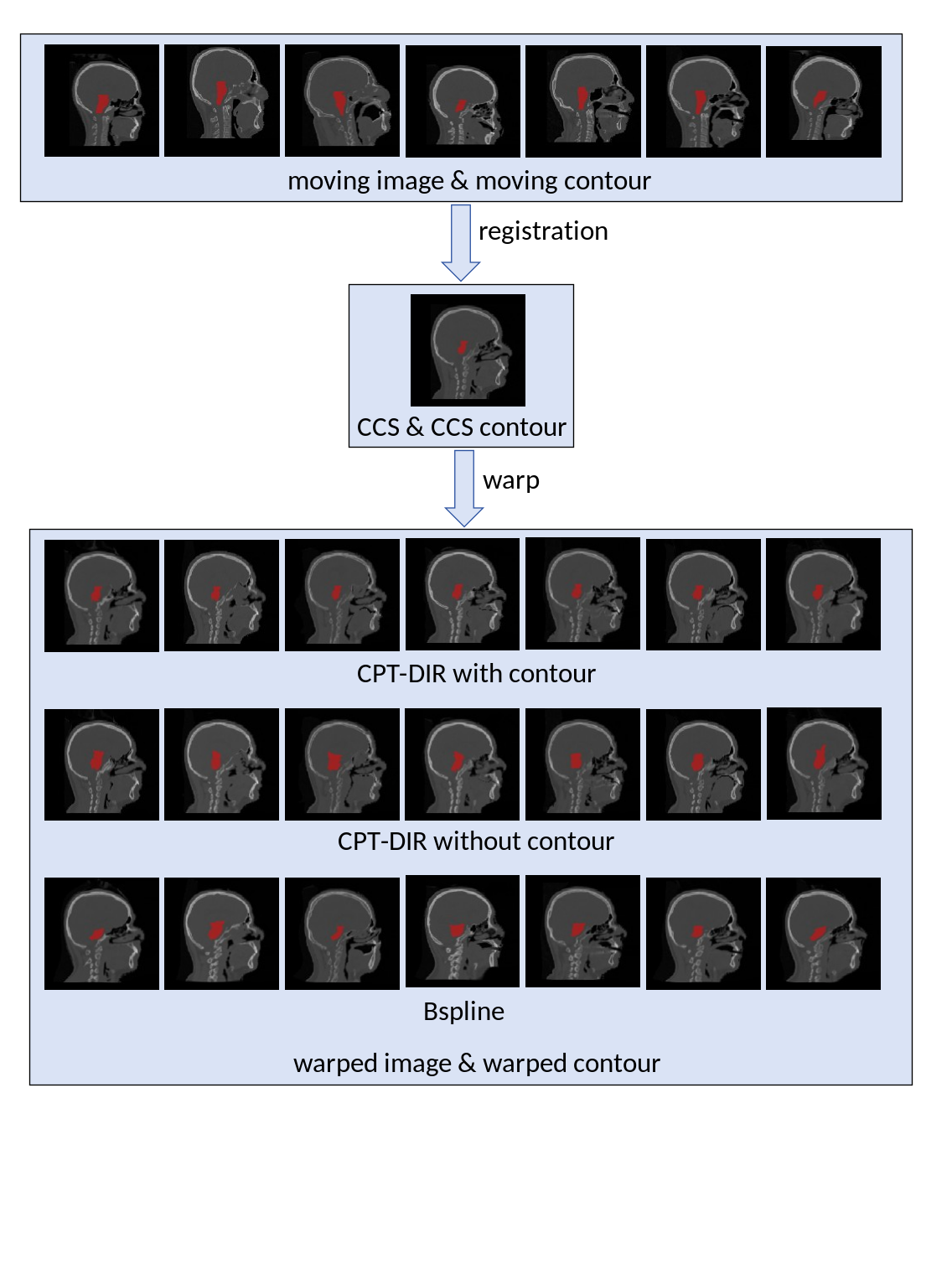}
\caption{Comparison of warped images and corresponding warped brainstem contours for CCS2.}
\label{fig:3}
\end{figure*}

The global registration performance of the two CCSs was quantitatively evaluated and summarised in Table~\ref{tab:1}. Across all metrics, the contour-informed CPT-DIR consistently outperformed the B-spline in both accuracy and stability. Notably, CPT-DIR reduced the MAE by 44.0\%, from 179.1 ± 17.8 HU to 98.9 ± 6.3 HU, and improved the SSIM by 1.3\%, from 0.974 ± 0.006 to 0.987 ± 0.002, and improved the PSNR by 22.6\%, from 21.292 ± 0.699 dB to 26.107 ± 0.636 dB. OAR contours and dose distributions of the seven patients with IMPT plans were warped to assess the benefit of including brainstem contours as an additional registration constraint, with the CCS1 resulting warped images and brainstem contours demonstrated in Figure~\ref{fig:2}, CCS2 warped images and warped brainstem contours demonstrated in Figure~\ref{fig:3}. It compared warped images and brainstem contours between the individual cases and the CCS, demonstrating that integrating the brainstem contour into the loss function significantly improved the DSC scores. Regarding the average results of two CCSs, compared to B-spline (0.604 ± 0.116) and CPT-DIR(wo) (0.621 ± 0.064), CPT-DIR(w) achieved a markedly higher brainstem DSC score (0.878 ± 0.017). For DOO metrics, CPT-DIR(w) (0.753 ± 0.043) also showed significant improvement over both B-spline (0.430 ± 0.117) and CPT-DIR(wo) (0.520 ± 0.085). For the warped brainstem, the HD95 was 2.31 ± 0.60 mm for CPT-DIR(w), compared to 8.98 ± 3.41 mm for CPT-DIR(wo) and 9.73 ± 3.94 mm for B-spline. Detailed HD95 values for the two CCSs were presented in Table~\ref{tab:2}.\\
\begin{table}[htbp]
\centering
\caption{ \textbf{Metrics comparison for different global DIR results}\\($\downarrow$: lower result means better, $\uparrow$: higher result means better)}
\tabcolsep=4.8mm
\begin{tabular}{lcc}
\toprule
 & \text{Method} & \text{Global image} \\
\midrule
                      & CPT-DIR(w)  & 98.9 ± 6.3 \\
MAE (HU) $\downarrow$ & CPT-DIR(wo) & 97.8 ± 7.4 \\
                      & B-Spline    & 179.1 ± 17.8 \\
\midrule
                            & CPT-DIR(w)  & 0.987 ± 0.002 \\
     SSIM (a.u.) $\uparrow$ & CPT-DIR(wo) & 0.987 ± 0.002 \\
                            & B-Spline    & 0.974 ± 0.006 \\
\midrule
                         & CPT-DIR(w)  & 26.107 ± 0.636 \\
   PSNR (dB) $\uparrow$  & CPT-DIR(wo) & 26.233 ± 0.722 \\
                         & B-Spline    & 21.292 ± 0.699 \\
\bottomrule
\end{tabular}
\label{tab:1}
\end{table}

\begin{table}[h]
\centering
\caption{\textbf{HD95 results}}
\tabcolsep=1.4mm
\begin{tabular}{*{4}{c}} 
\toprule
 & CPT-DIR(w) & CPT-DIR(wo) & B-spline \\
\midrule
CCS1 & 1.86 $\pm$ 0.20 & 7.26 $\pm$ 1.55 & 11.31 $\pm$ 4.46 \\
\midrule
CCS2 & 2.76 $\pm$ 0.53 & 10.69 $\pm$ 3.86 & 8.14 $\pm$ 2.48 \\
\midrule
CCS mean result & 2.31 $\pm$ 0.60 & 8.98 $\pm$ 3.41 & 9.73 $\pm$ 3.94 \\
\bottomrule
\end{tabular}
\label{tab:2}
\end{table}

\begin{figure*}[h]
\includegraphics
  [width=1\hsize]
  {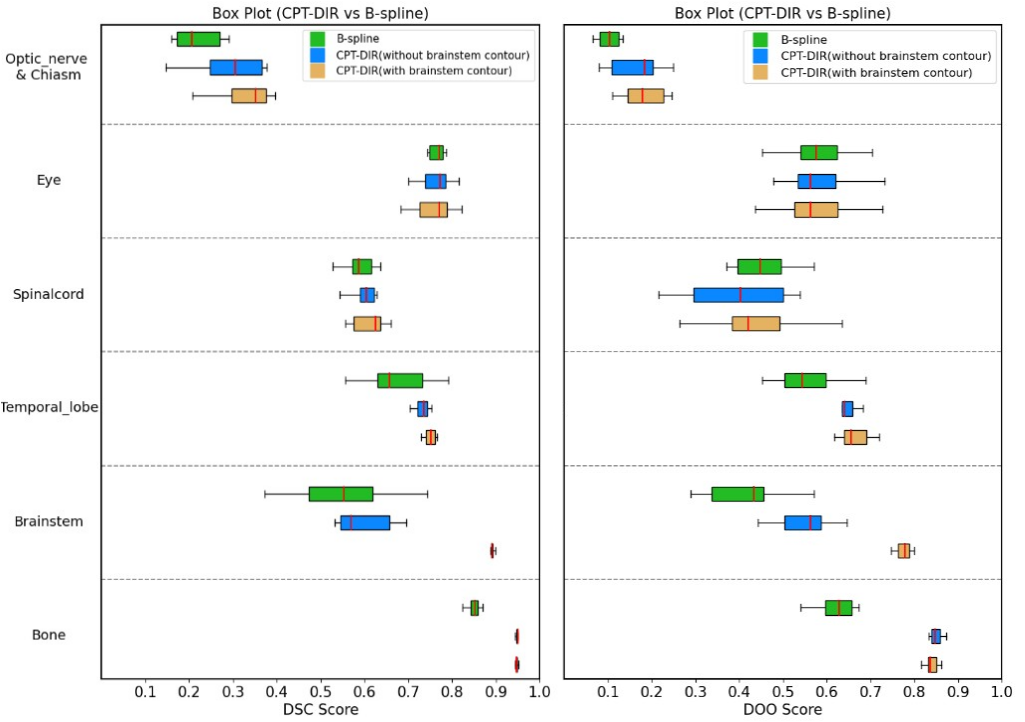}
\caption{Comparison of DSC and DOO scores across different DIR methods for CCS1.}
\label{fig:4}
\end{figure*}

\begin{figure*}[h]
\includegraphics
  [width=1\hsize]
  {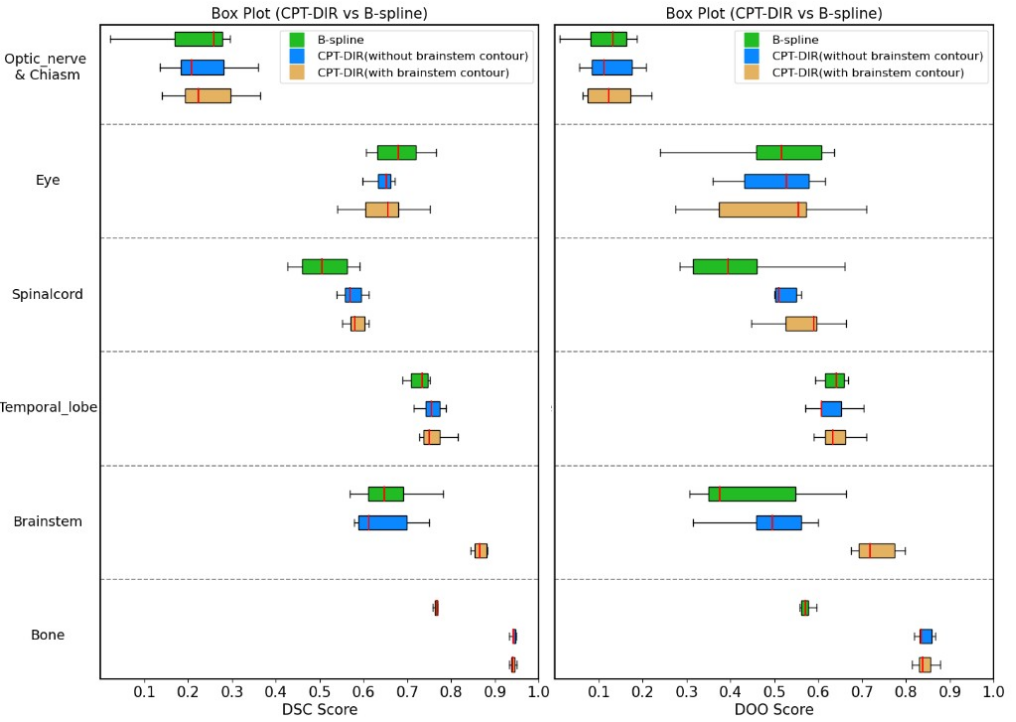}
\caption{Comparison of DSC and DOO scores across different DIR methods for CCS2.}
\label{fig:5}
\end{figure*}

\begin{figure*}[h]
\includegraphics
  [width=1\hsize]
  {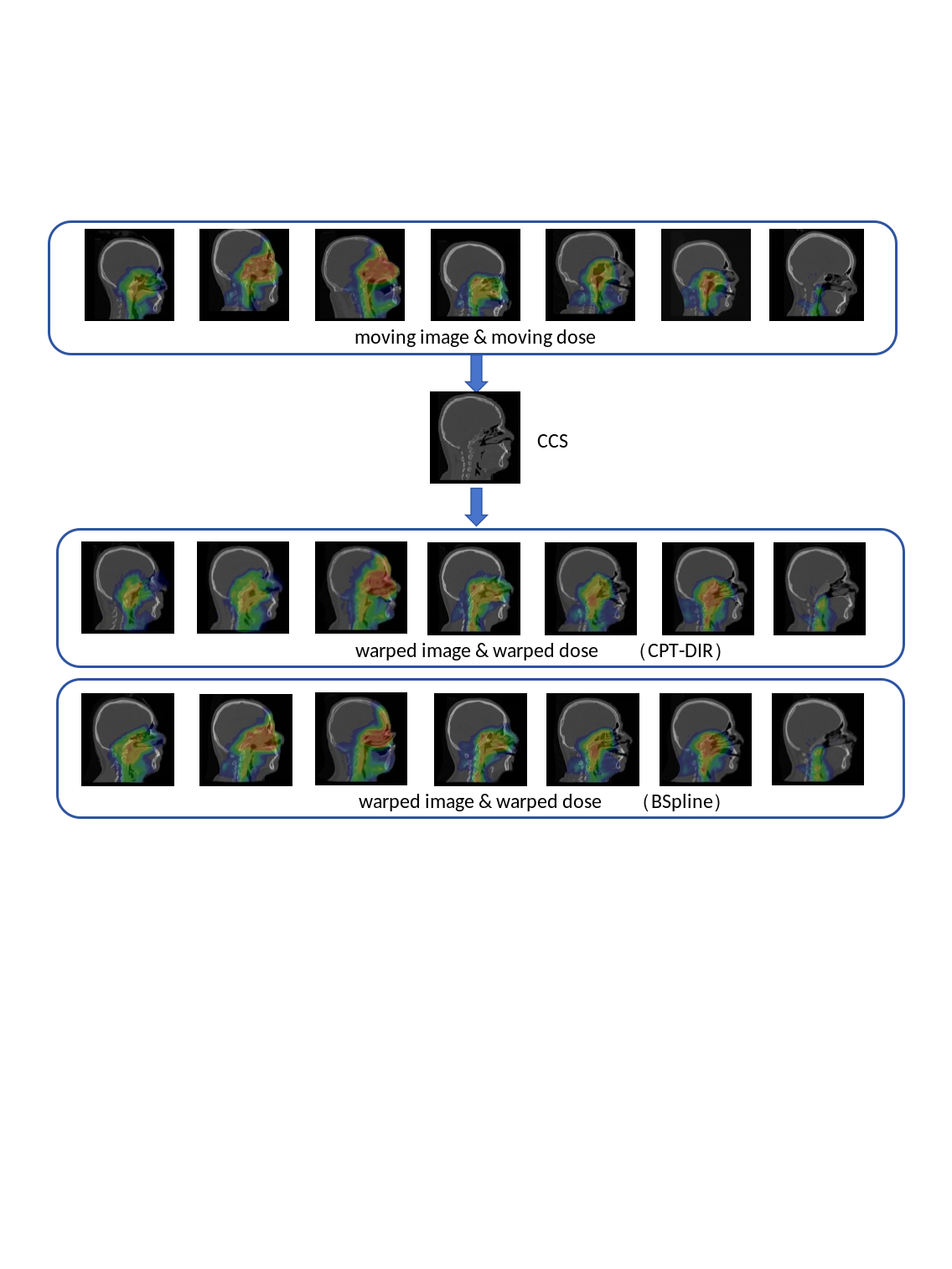}
\caption{Comparison of warped dose map and warped image of CCS2.}
\label{fig:7}
\end{figure*}

T-test statistical analysis demonstrated that CPT-DIR(w) produced superior outcomes compared to both B-spline and CPT-DIR(wo). The T-test statistical analysis results were summarised as Table~\ref{tab:3}.
Specifically for brainstem contours, the CPT-DIR(w) significantly increased DSC by 0.275 and DOO by 0.323 over B-spline (95\% CI: DSC [0.205, 0.345], p<0.001; DOO [0.256, 0.391], p<0.001). The CPT-DIR(wo) showed inconsistent performance, with DSC not significantly increased by 0.017 (95\% CI [-0.046, 0.080], p=0.2855) while DOO increased by 0.090 over B-spline (95\% CI [0.020, 0.160], p=0.0078). Direct comparison revealed the CPT-DIR(w) highly increased DSC by 0.258 and DOO by 0.233 over the CPT-DIR(wo) (95\% CI: DSC [0.217, 0.299], p<0.001; DOO [0.194, 0.273], p<0.001). 
Across all contours, CPT-DIR(w) increased DSC by 0.095 and DOO by 0.122 over B-spline (95\% CI: DSC [0.066, 0.123], p<0.001; DOO [0.086, 0.158], p<0.001). The CPT-DIR(wo) increased DSC by 0.044 and DOO by 0.073 over B-spline (95\% CI: DSC [0.022, 0.066], p<0.001; DOO [0.043, 0.102], p<0.001). Importantly, the CPT-DIR(w) increased DSC by 0.051 and DOO by 0.049 over CPT-DIR(wo) (95\% CI: DSC [0.028, 0.074], p<0.001; DOO [0.027, 0.071], p<0.001). \\
\hspace*{1em}The broader impact of the brainstem constraint on other OARs was assessed, with the DSC and DOO scores present in Figure~\ref{fig:4} and Figure~\ref{fig:5}, showing improved overall registration performance. All these findings highlighted the effectiveness of contour-informed regularisation in enhancing DIR accuracy, particularly in regions with poor soft tissue contrast, such as the skull. Incorporating the brainstem contour as an optimisation constraint in CPT-DIR significantly improved the DSC and DOO scores and reduces variance. More details for contour-specific results could be found in Table~\ref{tab:4}, Figure~\ref{fig:6} and  Figure~\ref{fig:7} showed warped dose distribution, which revealed key differences between results based on CPT-DIR and B-spline methods, despite dose comparisons were challenging due to large variability in this region.

\begin{table*}[htbp]
\centering
\caption{T-test statistical analysis results}
\tabcolsep=3.6mm
\begin{tabular}{ccccccc}  
\toprule
Region & Comparison Group & Metrics & Method & Improvement & 95\% CI & p-value \\
\midrule
\multirow{6}{*}{\centering Brainstem} 
 & \multirow{2}{*}{CPT-DIR(w) vs CPT-DIR(wo)} & DSC & CPT-DIR(w) & +0.258 & [0.217, 0.299] & <0.001 \\
 &                                            & DOO & CPT-DIR(w) & +0.233 & [0.194, 0.273] & <0.001 \\
\cline{2-7}
 & \multirow{2}{*}{CPT-DIR(w) vs B-spline}    & DSC & CPT-DIR(w) & +0.275 & [0.205, 0.345] & <0.001 \\
 &                                            & DOO & CPT-DIR(w) & +0.323 & [0.256, 0.391] & <0.001 \\
\cline{2-7}
 & \multirow{2}{*}{CPT-DIR(wo) vs B-spline}   & DSC & CPT-DIR(wo)& +0.017 & [-0.046, 0.080] & 0.2855 \\
 &                                            & DOO & CPT-DIR(wo)& +0.090 & [0.020, 0.160] & 0.0078 \\
\hline
\multirow{6}{*}{\centering All Contour} 
 & \multirow{2}{*}{CPT-DIR(w) vs CPT-DIR(wo)} & DSC & CPT-DIR(w) & +0.051 & [0.028, 0.074] & <0.001 \\
 &                                            & DOO & CPT-DIR(w) & +0.049 & [0.027, 0.071] & <0.001 \\
\cline{2-7}
 & \multirow{2}{*}{CPT-DIR(w) vs B-spline}    & DSC & CPT-DIR(w) & +0.095 & [0.066, 0.123] & <0.001 \\
 &                                            & DOO & CPT-DIR(w) & +0.122 & [0.086, 0.158] & <0.001 \\
\cline{2-7}
 & \multirow{2}{*}{CPT-DIR(wo) vs B-spline}   & DSC & CPT-DIR(wo)& +0.044 & [0.022, 0.066] & <0.001 \\
 &                                            & DOO & CPT-DIR(wo)& +0.073 & [0.043, 0.102] & <0.001 \\
\bottomrule
\end{tabular}
\label{tab:3}
\end{table*}

\begin{table*}[htb]
\centering
\caption{\textbf{Metrics comparison for organ DIR results}}
\tabcolsep=1.0mm
\begin{tabular}{l|ccccccc}
\toprule
\textbf{Metric} & \textbf{Method} & \textbf{Brainstem} & \textbf{Bone} & \textbf{Temporal lobe} & \textbf{Spinal cord} & \textbf{Eye} & \textbf{Optical nerve \& Chiasm} \\
\midrule
                & CPT-DIR(w)  & 0.891±0.004 & 0.946±0.004 & 0.750±0.013 & 0.611±0.039 & 0.758±0.049 & 0.329±0.064 \\
CCS1 Dice Score & CPT-DIR(wo) & 0.596±0.054 & 0.949±0.004 & 0.732±0.017 & 0.599±0.029 & 0.762±0.040 & 0.293±0.082 \\
                & B-Spline    & 0.550±0.122 & 0.850±0.014 & 0.673±0.079 & 0.588±0.036 & 0.751±0.045 & 0.254±0.119 \\
\midrule
                & CPT-DIR(w)  & 0.775±0.019 & 0.840±0.015 & 0.664±0.036 & 0.437±0.115 & 0.576±0.088 & 0.183±0.048 \\
CCS1 DOO Score  & CPT-DIR(wo) & 0.547±0.068 & 0.850±0.013 & 0.643±0.026 & 0.392±0.120 & 0.583±0.078 & 0.162±0.058 \\
                & B-Spline    & 0.411±0.091 & 0.621±0.044 & 0.555±0.078 & 0.455±0.069 & 0.579±0.077 & 0.137±0.103 \\
\midrule
                & CPT-DIR(w)  & 0.866±0.015 & 0.941±0.006 & 0.759±0.030 & 0.583±0.021 & 0.645±0.065 & 0.244±0.080 \\
CCS2 Dice Score & CPT-DIR(wo) & 0.645±0.065 & 0.942±0.005 & 0.755±0.025 & 0.574±0.026 & 0.654±0.041 & 0.233±0.078 \\
                & B-Spline    & 0.657±0.066 & 0.763±0.010 & 0.726±0.024 & 0.509±0.061 & 0.679±0.058 & 0.210±0.102 \\
\midrule
                & CPT-DIR(w)  & 0.732±0.046 & 0.843±0.020 & 0.641±0.039 & 0.566±0.070 & 0.490±0.143 & 0.129±0.060 \\
CCS2 DOO Score  & CPT-DIR(wo) & 0.493±0.091 & 0.843±0.018 & 0.627±0.044 & 0.527±0.072 & 0.503±0.089 & 0.128±0.057 \\
                & B-Spline    & 0.449±0.128 & 0.566±0.022 & 0.636±0.027 & 0.418±0.127 & 0.504±0.130 & 0.117±0.063 \\
\bottomrule
\end{tabular}
\label{tab:4}
\end{table*}

  \section{Discussion}
\begin{figure*}[h]
\includegraphics
  [width=1\hsize]
  {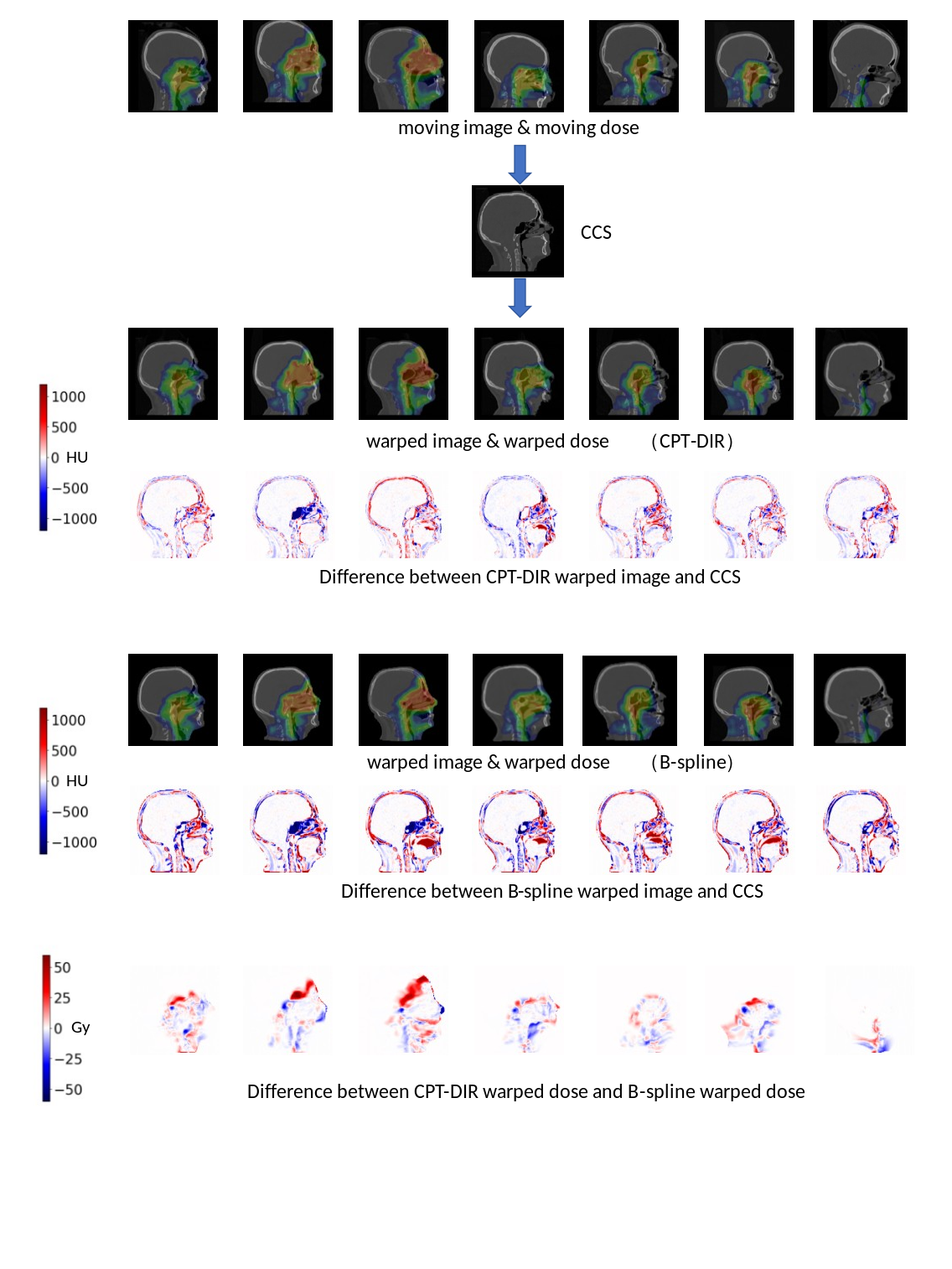}
\caption{DIR-based warping results: image and dose distribution comparisons for CCS1.}
\label{fig:6}
\end{figure*}

This study demonstrated an advanced HN image registration approach using updated CPT-DIR with incorporated contour information. Contours were included in the loss function as constraints to improve the local alignment of corresponding structures in the registered image. This approach not only enhanced the overall accuracy of image registration but also improved the consistency of target region alignment for VBA.\\
\hspace*{1em}Accurate spatial normalisation was essential for VBA, as it directly influenced the reliability of analyses linked to clinical outcomes~\cite{bib:24,bib:25}. The proposed method enabled more precise spatial normalisation in regions requiring detailed evaluation, potentially improving the effectiveness of VBA in assessing treatment-related toxicities in future applications. In VBA workflows, inter-patient image registration was a critical yet challenging step, particularly due to substantial anatomical variability across patients. Systematic evaluation of DIR accuracy was therefore essential, as registration errors could impact subsequent analysis~\cite{bib:14,bib:26,bib:27,bib:28}. Figure~\ref{fig:6} also illustrated differences in image registration and warped dose distributions between the contour-informed CPT-DIR and B-spline methods relative to the CCS. The image difference revealed that CPT-DIR with contour constraints produced lower HU errors, particularly at tissue and bone boundaries. Alongside improved DSC and DOO scores, these results suggested that the proposed method offered a more robust and accurate registration approach for VBA. Although, in this study, only whole-organ contours were used as constraints, future work may explore combining multiple OAR contours or using organ substructures to refine constraint design further. Additionally, as shown by the dose differences in Figure~\ref{fig:6}, the choice of registration method significantly influenced dose normalisation, potentially affecting the identification of voxel-level predictors of treatment-related toxicity. Understanding and addressing these variations will be critical in advancing VBA outcomes' reliability and clinical relevance.\\
\hspace*{1em}Using DVF for dose spatial normalisation was one of the most critical steps in VBA~\cite{bib:6,bib:29}. Although our upgraded approach achieves higher accuracy in evaluation metrics, the warped dose distribution derived by the original CPT-DIR appear less smooth. This difference arose from the fundamental design of the two methods. B-spline generated the DVF by interpolating displacements at control points arranged on a predefined grid, inherently promoting smoothness due to the continuity of spline functions. In contrast, using an implicit neural network representation, CPT-DIR predicted the DVF at each voxel. While smoothness was encouraged through regularisation terms in the loss function, the voxel-wise nature of CPT-DIR lacks the explicit and strong smoothness constraints present in B-spline interpolation. As a result, CPT-DIR may produce sharper transitions and less smooth, warped dose distributions, while achieving superior accuracy according to standard evaluation metrics. Notably, CPT-DIR did not require external datasets for network training. The network took pixel coordinates from fixed and moving images as input and output the velocity vector field (VVF). This approach overcame the limitation of medical image accessibility, as most clinical data could not be used for research due to privacy concerns~\cite{bib:30}. Our method introduced an innovative registration framework that eliminated the dependency on external training datasets while optimizing contour-based DIR.\\
\hspace*{1em}The selection of registration parameters significantly impacted the registration performance~\cite{bib:31,bib:32,bib:33}. Theoretically, incorporating more contour constraints could further improve registration accuracy; however, practically, achieving stable convergence required careful balancing of the weighting coefficient $\alpha$ in the target loss function. Selecting inappropriate weighting coefficients could lead to network divergence during training, causing model instability and significant registration errors. Overemphasising particular contours, by assigning excessively high weights to OARs or specific regions, could overly constrain the registration in those areas, ultimately compromising accuracy in other regions. Balancing these factors was, therefore, essential for robust and generalisable registration performance.\\
\hspace*{1em}From the result of Table~\ref{tab:1} and Table~\ref{tab:4}, contour constraints could not comprehensively improve the global registration result of CPT-DIR, but could improve the registration accuracy of corresponding contour. As shown in Table~\ref{tab:4}, smaller structures demonstrated both lower DSC scores and DOO metrics, attributable to the inherent difficulty in achieving accurate registration for small-volume anatomical features. In t-test statistical analysis result of the improved DSC and DOO metrics results of warped brainstem, CPT-DIR(wo) showed different P-values. Because the DSC parameter was calculated by binary masks, which ignored the differences caused by voxel dose deformation. The DOO metric intrinsically accounted for per-voxel dose variations, making it sensitive to deformations affecting local dose distribution. The same result conclusion was also presented in Figure~\ref{fig:4} and Figure~\ref{fig:5}. Additionally, contour constraints could indeed significantly improve the alignment of corresponding structures in the registered image from the t-test statistical analysis result.\\
\hspace*{1em}In this study, the dataset available was relatively small. Nevertheless, this sample size (=7) was sufficient to demonstrate the proof of concept for proposed DIR algorithm, given the better global registration result and regional registration accuracy. Only two patient images were randomly selected as the CCS in this study. Although results from both cases showed that incorporating contour constraints improved contour consistency, the limited number of CCS cases might affect the robustness of the analysis. However, prior research indicated that increasing the number of CCSs did not necessarily improve registration accuracy~\cite{bib:15}, and the choice of reference patient did not impact the identified regions~\cite{bib:34}. To enhance robustness and explore the full potential of the proposed approach, future work will involve using multiple CCS images and incorporating registration based on HN MRI and MRI-derived contours for further evaluation.\\
\hspace*{1em}This study proposed an optimised registration method based on CPT-DIR with integrated contour constraints for more precise HN image registration. By directly incorporating contour information into the target loss function, the method improved global registration accuracy while enhancing consistency within target regions which were clinically critical areas requiring more precise alignment. This approach enabled more precise spatial normalisation and offered a novel solution for advancing spatial normalisation in VBA research and applications.

 \section{Acknowledgement}
 This research was supported by the project “Increased Precision for Personalized Cancer Treatment Delivery Utilizing 4D Adapted Proton Therapy (EPIC-4DAPT),” funded by the Swiss National Science Foundation (SNSF) under grant number 212855. The data registration tools utilized in this work were developed with support from the Personalized Health and Related Technologies (PHRT) as part of the interdisciplinary doctoral grant (iDoc 2021-360) of the ETH domain, Switzerland. 

\begin{thebibliography}{99}

\bibitem{bib:1} {Palma G, Monti S, Cella L. Voxel-based analysis in radiation oncology: A methodological cookbook. Phys Med 2020; 69:192-204.}
\href{https://doi.org/10.1016/j.ejmp.2019.12.013}{https://doi.org/10.1016/j.ejmp.2019.12.013}

\bibitem{bib:2}
{Shipley WU, Tepper JE, Prout GR, Verhey LJ, Mendiondo OA, Goitein M, Koehler AM, Suit HD. Proton radiation as boost therapy for localized prostatic carcinoma. Jama 1979; 241:1912-15.}
\href{https://doi.org/10.1001/jama.1979.03290440034024}{https://doi.org/10.1001/jama.1979.03290440034024}

\bibitem{bib:3}
{Tommasino F, Nahum A, Cella L. Increasing the power of tumour control and normal tissue complication probability modelling in radiotherapy: recent trends and current issues. Transl Cancer Res 2017; S807-S21.}
\href{https://doi.org/10.21037/tcr.2017.06.0}{https://doi.org/10.21037/tcr.2017.06.0}

\bibitem{bib:4}
{Ghita M, Gill EK, Walls GM, Edgar KS, Mcmahon SJ, Osorio EV, Bergom C, Grieve DJ, Watson CJ, Mcwilliam A. Cardiac sub-volume targeting demonstrates regional radiosensitivity in the mouse heart. Radiother Oncol 2020; 152:216-21.}
\href{https://doi.org/10.1016/j.radonc.2020.07.016}{https://doi.org/10.1016/j.radonc.2020.07.016}

\bibitem{bib:5}
{Monti S, Palma G, D’avino V, Gerardi M, Marvaso G, Ciardo D, Pacelli R, Jereczek-Fossa BA, Alterio D, Cella L. Voxel-based analysis unveils regional dose differences associated with radiation-induced morbidity in head and neck cancer patients. Sci Rep 2017; 7:7220.}
\href{https://doi.org/10.1038/s41598-017-07586-x}{https://doi.org/10.1038/s41598-017-07586-x}

\bibitem{bib:6}
{Mcwilliam A, Palma G, Abravan A, Acosta O, Appelt A, Aznar M, Monti S, Onjukka E, Panettieri V, Placidi L. Voxel-based analysis: roadmap for clinical translation. Radiother Oncol 2023; 109868.}
\href{https://doi.org/10.1016/j.radonc.2023.109868}{https://doi.org/10.1016/j.radonc.2023.109868}

\bibitem{bib:7}
{Spampinato S, Fokdal L, Marinovskij E, Axelsen S, Pedersen E, Pötter R, Lindegaard J, Tanderup K. Assessment of dose to functional sub-structures in the lower urinary tract in locally advanced cervical cancer radiotherapy. Phys Med 2019; 59:127-32.}
\href{https://doi.org/10.1016/j.ejmp.2019.01.017}{https://doi.org/10.1016/j.ejmp.2019.01.017}

\bibitem{bib:8}
{Jaikuna T, Osorio EV, Azria D, Chang-Claude J, De Santis MC, Gutiérrez-Enríquez S, Van Herk M, Hoskin P, Lambrecht M, Lingard Z. Contouring variation affects estimates of normal tissue complication probability for breast fibrosis after radiotherapy. The Breast 2023; 72:103578.}
\href{https://doi.org/10.1016/j.breast.2023.103578}{https://doi.org/10.1016/j.breast.2023.103578}

\bibitem{bib:9}
{Ebert MA, Gulliford S, Acosta O, De Crevoisier R, Mcnutt T, Heemsbergen WD, Witte M, Palma G, Rancati T, Fiorino C. Spatial descriptions of radiotherapy dose: normal tissue complication models and statistical associations. Physics in Medicine \& Biology 2021; 66:12TR01.}
\href{https://doi.org/10.1088/1361-6560/ac0681}{https://doi.org/10.1088/1361-6560/ac0681}

\bibitem{bib:10}
{Beasley W, Thor M, Mcwilliam A, Green A, Mackay R, Slevin N, Olsson C, Pettersson N, Finizia C, Estilo C. Image-based data mining to probe dosimetric correlates of radiation-induced trismus. Int J Radiat Oncol Biol Phys 2018; 102:1330-38.}
\href{https://doi.org/10.1016/j.ijrobp.2018.05.054}{https://doi.org/10.1016/j.ijrobp.2018.05.054}

\bibitem{bib:11}
{Mylona E, Cicchetti A, Rancati T, Palorini F, Fiorino C, Supiot S, Magne N, Crehange G, Valdagni R, Acosta O. Local dose analysis to predict acute and late urinary toxicities after prostate cancer radiotherapy: Assessment of cohort and method effects. Radiother Oncol 2020; 147:40-49.}
\href{https://doi.org/10.1016/j.radonc.2020.02.028}{https://doi.org/10.1016/j.radonc.2020.02.028}

\bibitem{bib:12}
{Monti S, Xu T, Mohan R, Liao Z, Palma G, Cella L. Radiation-induced esophagitis in non-small-cell lung cancer patients: voxel-based analysis and NTCP modeling. Cancers 2022; 14:1833.}
\href{https://doi.org/10.3390/cancers14071833}{https://doi.org/10.3390/cancers14071833}

\bibitem{bib:13}
{Monti S, Xu T, Liao Z, Mohan R, Cella L, Palma G. On the interplay between dosiomics and genomics in radiation-induced lymphopenia of lung cancer patients. Radiother Oncol 2022; 167:219-25.}
\href{https://doi.org/10.1016/j.radonc.2021.12.038}{https://doi.org/10.1016/j.radonc.2021.12.038}

\bibitem{bib:14}
{Rigaud B, Simon A, Castelli J, Lafond C, Acosta O, Haigron P, Cazoulat G, De Crevoisier R. Deformable image registration for radiation therapy: principle, methods, applications and evaluation. Acta Oncol 2019; 58:1225-37.}
\href{https://doi.org/10.1080/0284186X.2019.1620331}{https://doi.org/10.1080/0284186X.2019.1620331}

\bibitem{bib:15}
{Jaikuna T, Wilson F, Azria D, Chang-Claude J, De Santis MC, Gutiérrez-Enríquez S, Van Herk M, Hoskin P, Kotzki L, Lambrecht M. Optimising inter-patient image registration for image-based data mining in breast radiotherapy. Phys Imaging Radiat Oncol 2024; 32:100635.}
\href{https://doi.org/10.1016/j.phro.2024.100635}{https://doi.org/10.1016/j.phro.2024.100635}

\bibitem{bib:16}
{Rueckert D, Sonoda LI, Hayes C, Hill DL, Leach MO, Hawkes DJ. Nonrigid registration using free-form deformations: application to breast MR images. IEEE TMI 1999; 18:712-21.}
\href{https://doi.org/10.1109/42.796284}{https://doi.org/10.1109/42.796284}

\bibitem{bib:17}
{Wei D, Li Y-M. Generalized sampling expansions with multiple sampling rates for lowpass and bandpass signals in the fractional Fourier transform domain. IEEE Trans Signal Process 2016; 64:4861-74.}
\href{https://doi.org/10.1109/TSP.2016.2560148}{https://doi.org/10.1109/TSP.2016.2560148}

\bibitem{bib:18}
{Rueckert D, Aljabar P, Heckemann RA, Hajnal JV, Hammers A. Diffeomorphic registration using B-splines; proceedings of the Medical Image Computing and Computer-Assisted Intervention–MICCAI 2006,Part II 9, 2006.}

\bibitem{bib:19}
{Amstutz F, D’almeida PG, Wu X, Albertini F, Bachtiary B, Weber DC, Unkelbach J, Lomax AJ, Zhang Y. Quantification of deformable image registration uncertainties for dose accumulation on head and neck cancer proton treatments. Phys Med 2024; 122:103386.}
\href{https://doi.org/10.1016/j.ejmp.2024.103386}{https://doi.org/10.1016/j.ejmp.2024.103386}

\bibitem{bib:20}
{Li X, Li M, Lomax A, Buhmann J, Zhang Y. Continuous sPatial-Temporal Deformable Image Registration (CPT-DIR) for motion modelling in radiotherapy: beyond classic voxel-based methods. arXiv preprint 2024.}
\href{https://doi.org/10.48550/arXiv.2405.15385}{https://doi.org/10.48550/arXiv.2405.15385}

\bibitem{bib:21}
{Wasserthal J, Breit H-C, Meyer MT, Pradella M, Hinck D, Sauter AW, Heye T, Boll DT, Cyriac J, Yang S. TotalSegmentator: robust segmentation of 104 anatomic structures in CT images. Radiol Artif Intell 2023; 5.}
\href{https://doi.org/10.1148/ryai.230024}{https://doi.org/10.1148/ryai.230024}

\bibitem{bib:22}
{Dréan G, Acosta O, Lafond C, Simon A, De Crevoisier R, Haigron P. Interindividual registration and dose mapping for voxelwise population analysis of rectal toxicity in prostate cancer radiotherapy. Med Phys 2016; 43:2721-30.}
\href{https://doi.org/10.1118/1.4948501}{https://doi.org/10.1118/1.4948501}

\bibitem{bib:23}
{Sharp GC, Li R, Wolfgang J, Chen G, Peroni M, Spadea MF, Mori S, Zhang J, Shackleford J, Kandasamy N. Plastimatch: an open source software suite for radiotherapy image processing; proceedings of the Proceedings of the XVI’th International Conference on the use of Computers in Radiotherapy (ICCR), 2010.}

\bibitem{bib:24}
{Cella L, Monti S, Xu T, Liuzzi R, Stanzione A, Durante M, Mohan R, Liao Z, Palma G. Probing thoracic dose patterns associated to pericardial effusion and mortality in patients treated with photons and protons for locally advanced non-small-cell lung cancer. Radiother Oncol 2021; 160:148-58.}
\href{https://doi.org/10.1016/j.radonc.2021.04.025}{https://doi.org/10.1016/j.radonc.2021.04.025}

\bibitem{bib:25}
{Walls GM, Ghita M, Queen R, Edgar KS, Gill EK, Kuburas R, Grieve DJ, Watson CJ, Mcwilliam A, Van Herk M. Spatial gene expression changes in the mouse heart after base-targeted irradiation. Int J Radiat Oncol Biol Phys 2023; 115:453-63.}
\href{https://doi.org/10.1016/j.ijrobp.2022.08.031}{https://doi.org/10.1016/j.ijrobp.2022.08.031}

\bibitem{bib:26}
{Brock KK, Mutic S, Mcnutt TR, Li H, Kessler ML. Use of image registration and fusion algorithms and techniques in radiotherapy: Report of the AAPM Radiation Therapy Committee Task Group No. 132. Medical physics 2017; 44:e43-e76.}
\href{https://doi.org/10.1002/mp.12256}{https://doi.org/10.1002/mp.12256}

\bibitem{bib:27}
{Palma G, Monti S, Xu T, Scifoni E, Yang P, Hahn SM, Durante M, Mohan R, Liao Z, Cella L. Spatial dose patterns associated with radiation pneumonitis in a randomized trial comparing intensity-modulated photon therapy with passive scattering proton therapy for locally advanced non-small cell lung cancer. International Journal of Radiation Oncology* Biology* Physics 2019; 104:1124-32.}
\href{https://doi.org/10.1016/j.ijrobp.2019.02.039}{https://doi.org/10.1016/j.ijrobp.2019.02.039}

\bibitem{bib:28}
{Rong Y, Rosu-Bubulac M, Benedict SH, Cui Y, Ruo R, Connell T, Kashani R, Latifi K, Chen Q, Geng H. Rigid and deformable image registration for radiation therapy: a self-study evaluation guide for NRG oncology clinical trial participation. Practical radiation oncology 2021; 11:282-98.}
\href{https://doi.org/10.1016/j.prro.2021.02.007}{https://doi.org/10.1016/j.prro.2021.02.007}

\bibitem{bib:29}
{Murr M, Brock KK, Fusella M, Hardcastle N, Hussein M, Jameson MG, Wahlstedt I, Yuen J, Mcclelland JR, Osorio EV. Applicability and usage of dose mapping/accumulation in radiotherapy. Radiother Oncol 2023; 182:109527.}
\href{https://doi.org/10.1016/j.radonc.2023.109527}{https://doi.org/10.1016/j.rad onc.2023.109527}

\bibitem{bib:30}
{Dhar T, Dey N, Borra S, Sherratt RS. Challenges of deep learning in medical image analysis—improving explainability and trust. IEEE Transactions on Technology and Society 2023; 4:68-75.}


\bibitem{bib:31}
{Chetty IJ, Rosu-Bubulac M. Deformable registration for dose accumulation; proceedings of the Seminars in Radiation Oncology, 2019. Elsevier.}


\bibitem{bib:32}
{Crum WR, Hartkens T, Hill D. Non-rigid image registration: theory and practice. The British journal of radiology 2004; 77:S140-S53.}
\href{https://doi.org/10.1259/bjr/25329214}{https://doi.org/10.1259/bjr/25329214}

\bibitem{bib:33}
{Schultheiss TE, Tome WA, Orton CG. Point/counterpoint: it is not appropriate to "deform" dose along with deformable image registration in adaptive radiotherapy. Med Phys 2012; 39:6531-3.}
\href{https://doi.org/10.1118/1.4722968}{https://doi.org/10.1118/1.4722968}

\bibitem{bib:34}
{Osorio EV. Estimating inter-patient registration uncertainty for image-based data mining; proceedings of the Biomedical Image Registration: 8th International Workshop, 2018.}

\end{thebibliography}
\end{document}